\documentclass[trackchanges,twocolumn]{aastex7}
\usepackage{amsmath}
\usepackage{appendix}

\begin{document}

\title{Radio Study of Vela X Cocoon}

\author[sname='Liu', gname='Yihan']{Yihan Liu}
\altaffiliation{These authors contribute equally to this work.}
\affiliation{School of Physics and Astronomy, Sun Yat-Sen University, No. 2 Daxue Road, 519082, Zhuhai China}
\email[show]{liuyh363@mail.sysu.edu.cn}

\author[sname='Zhang', gname='Yu']{Yu Zhang}
\altaffiliation{These authors contribute equally to this work.}
\affiliation{School of Physics and Astronomy, Sun Yat-Sen University, No. 2 Daxue Road, 519082, Zhuhai China}
\email{zhangy969@mail2.sysu.edu.cn}

\author[0000-0002-5847-2612,sname='Ng', gname='Stephen Chi-Yung']{C.-Y. Ng}
\affiliation{Department of Physics, The University of Hong Kong, Pokfulam Road, Hong Kong}
\affiliation{Hong Kong Institute for Astronomy and Astrophysics, The University of Hong Kong, Pokfulam Road, Hong Kong}
\email{ncy@astro.physics.hku.hk}

\author[sname='Qiu', gname='Zijian']{Zijian Qiu}
\affiliation{School of Physics and Astronomy, Sun Yat-Sen University, No. 2 Daxue Road, 519082, Zhuhai China}
\email{qiuzj3@mail2.sysu.edu.cn}

\author[sname='Lin', gname='Sujie']{Sujie Lin}
\affiliation{School of Physics and Astronomy, Sun Yat-Sen University, No. 2 Daxue Road, 519082, Zhuhai China}
\email{linsj6@mail.sysu.edu.cn}

\author[0000-0001-7416-7434, sname='Yang', gname='Lili']{Lili Yang}
\affiliation{School of Physics and Astronomy, Sun Yat-Sen University, No. 2 Daxue Road, 519082, Zhuhai China}
\affiliation{Centre for Astro-Particle Physics, University of Johannesburg, P.O. Box 524, Auckland Park 2006, South Africa}
\email{yanglli5@mail.sysu.edu.cn}


\begin{abstract}
The evolution of pulsar Wind Nebulae (PWNe) influences how high energy particles in the vicinity are generated and transport. 
The Vela PWN (only $\sim300$\,pc away), provides a rather rare case between young and well-evolved systems.
We therefore performed new 6 and 16\,cm high-resolution observations of the Vela X Cocoon region with the Australia Telescope Compact Array (ATCA). 
The observations reveal a complex region with a $\sim0.5^\circ$ major curved filament extending to far south from the pulsar, as well as other intersecting filaments and wisps. Our spectral analysis hints its connection with the PWN. 
Our results also found strongly linearly polarized emission, ordered and tangential $B$-field to the filaments.
We find the rotation measure (RM) and polarization fraction (PF) along the filament are anti-correlated with the total intensity. 
We develop a simple 3D model of a spiral filament to explain these,
while the PF distribution requires external interpretations such as interaction with the reverse shock. 
Comparison with archival data suggests that large scale features like the major filament are generally stable and large motions near the X-ray filament, all these confirm the distinction between radio and X-ray features.
\end{abstract}


\section{Introduction}
Pulsars are neutron stars left behind by supernova explosions, known for stable rotation and periodic pulsations. 
For some energetic pulsars, out-flowing pulsar winds inflate high energy magnetized bubbles in the vicinity, relativistic particles inside generate emissions from radio up to very high energies (over TeV), observed as pulsar wind nebulae (PWNe).
PWNe are important galactic high energy sources, many mysterious TeV sources in the first LHAASO catalog are spatially associated with a pulsar or PWN \citep{2024ApJS..271...25C}, while it is yet to be well illustrated how PWN high energy particles transport and connect with these TeV sources; for example, it is common to see offsets between TeV emissions and pulsars/PWNe \citep[see][]{2008ICRC....3.1341W,2024ApJS..271...25C}.

Multi-wavelength observations PWN trace these high energy particles.
Many young PWNe have geometries of an equatorial torus/tori with jets in polar regions in X-rays \citep{2004ApJ...601..479N,2008ApJ...673..411N}, which could be related to freshly injected relativistic particles close to the termination shock (TS).
When propagating outward, relativistic particles experience significant synchrotron (SYN) cooling and become invisible, radio emissions (less affected by cooling) reflect integrated history over PWN evolution and local $B$-field properties.
Most young radio PWNe are however so misaligned with X-ray structures (except the PWN G11.2-0.3), with even more diverse morphologies than rather simple and unified X-ray nebular geometry \citep{2017ApJ...840...82D,2025ApJ...988..163Z,2006ApJ...645.1180B}.
For well evolved PWN systems, the pulsar eventually exits the SNR and travels supersonically in the interstellar medium (ISM), a compact and bow-shocked head close to the pulsar in multi-wavelength is expected in this stage; besides, long trails behind the pulsar shows in radio and X-ray observations \citep{2008AIPC..983..171K,2010ApJ...712..596N}.
All these implies that PWN evolution influences particle transport inside.

The Vela PWN-SNR system is a rare case connecting very young and well evolved systems, it is regarded that the PWN has just experienced interactions with SNR ejecta.
Vela SNR has a radio shell with a huge diameter of $\sim6^\circ$ \citep{1968AuJPh..21..201M}, hosting a pulsar (PSR~B0833-45) with a spin period of 89.3\,ms, characteristic age $\tau_c$=11.3\,kyr, spin down power $\dot{E}=7\times 10^{36}$\,erg/s \citep{dml02}, and is only $\sim$290\,pc away \citep{2003ApJ...596.1137D}.
This pulsar also powers a jet/torus X-ray PWN system \citep{2001ApJ...556..380H}, the radio toroidal nebula is also observed \citep{2003ApJ...596.1137D}.
Intriguingly beyond the central region, the PWN extends so asymmetrically to the south more than 1$^\circ$ away (the Vela X ``Cocoon" region). 
Its SYN emissions are visible in many wavelengths, for example, radio and X-rays, but interestingly distinct with each other \citep{2018ApJ...865...86S}; besides, TeV gamma-ray emissions with similar elongated geometry in the HESS observation also suggests that particles inside are so energetic \citep{2006A&A...448L..43A}.
Though it is not conclusive how the Vela X ``Cocoon" is formed, observations hints its possible relationship with the anisotropic SNR environment \citep{1998AJ....116.1886B,1999A&A...342..839B}, that reverse shock (RS) back from the north may have collided and disrupted the original PWN \citep[see][]{2004A&A...420..937V,2015ApJ...808..100T,2018ApJ...865...86S}.
Their hydrodynamic models after a proposed interaction is generally aligned with the morphology of the Vela PWN-SNR system from observations, while ignored the contribution of $B$-field \citep{2018ApJ...865...86S}.
Nonetheless, much about the Vela X ``Cocoon" are puzzling, for example, why the radio and X-ray ``Cocoon" are so separated and why there are only significant radio and X-ray ``Cocoon" features far away (beyond $\sim10\arcmin$) from the pulsar.

Though there were previous data of Vela X Cocoon in radio bands \citep{1995MNRAS.277.1435M}, very few has systematic full polarization and high resolution observations in different radio frequencies. Therefore, this study performed high resolution new observations with the Australia Telescope Compact Array (ATCA) at different frequencies, and intends to further study the Vela X PWN morphology, polarization (PL), and $B$-field properties, which could be meaningful for further magneto-hydrodynamical simulations. 
Our observations and data reduction processes will be briefly introduced in Section \ref{sec:obs_data}; our main results are in Section \ref{sec:result}.
Then we will discuss and conclude this study in Section \ref{sec:discussion} and \ref{sec:conclusion}, respectively. 
\section{Observations and Data Reduction}
\label{sec:obs_data}
\begin{table}[b!]
\label{tab:radio_obs}
    \footnotesize
    \centering
    \begin{tabular}{ccccc}
    \toprule
       Obs   &  Array   &  Center      &  Bandwidth &  Integration \\
       Date  &  Config. &  Freq. (MHz) &  (MHz)     &  Time (hr)   \\
       \hline
        2024 Nov 23 &  750D  & 2100 & 2048 & 9.6 \\
        2024 Nov 24 &  750D  & 5500 & 2048 & 10.2 \\
        2024 Nov 25 &  750D  & 5500 & 2048 & 9.5 \\
         \hline
    \end{tabular}
    \caption{ATCA observations in this study}
    \label{tab:radioobs}
\end{table}

\begin{figure*}[ht!]
    \centering
    \includegraphics[width=0.49\linewidth]{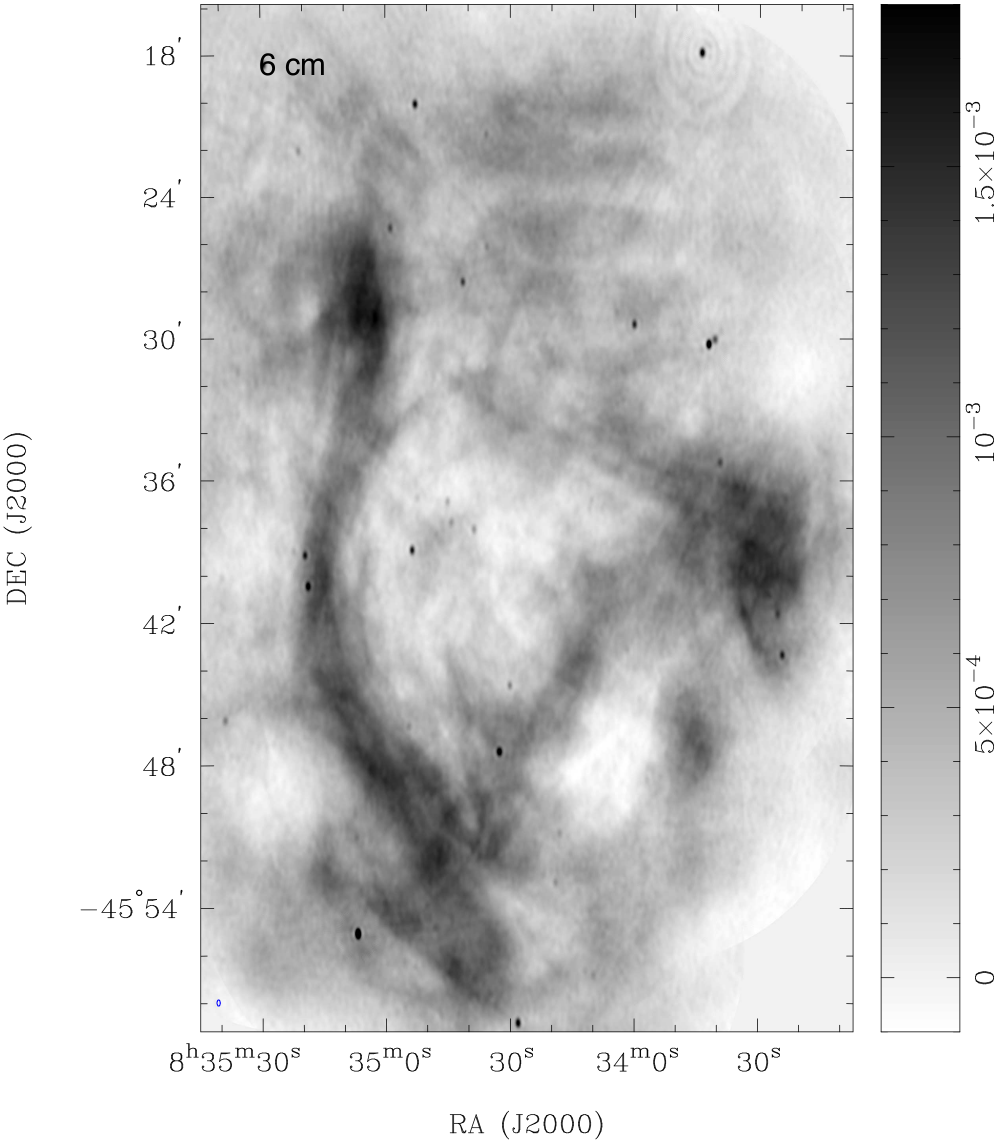}
    \includegraphics[width=0.49\linewidth]{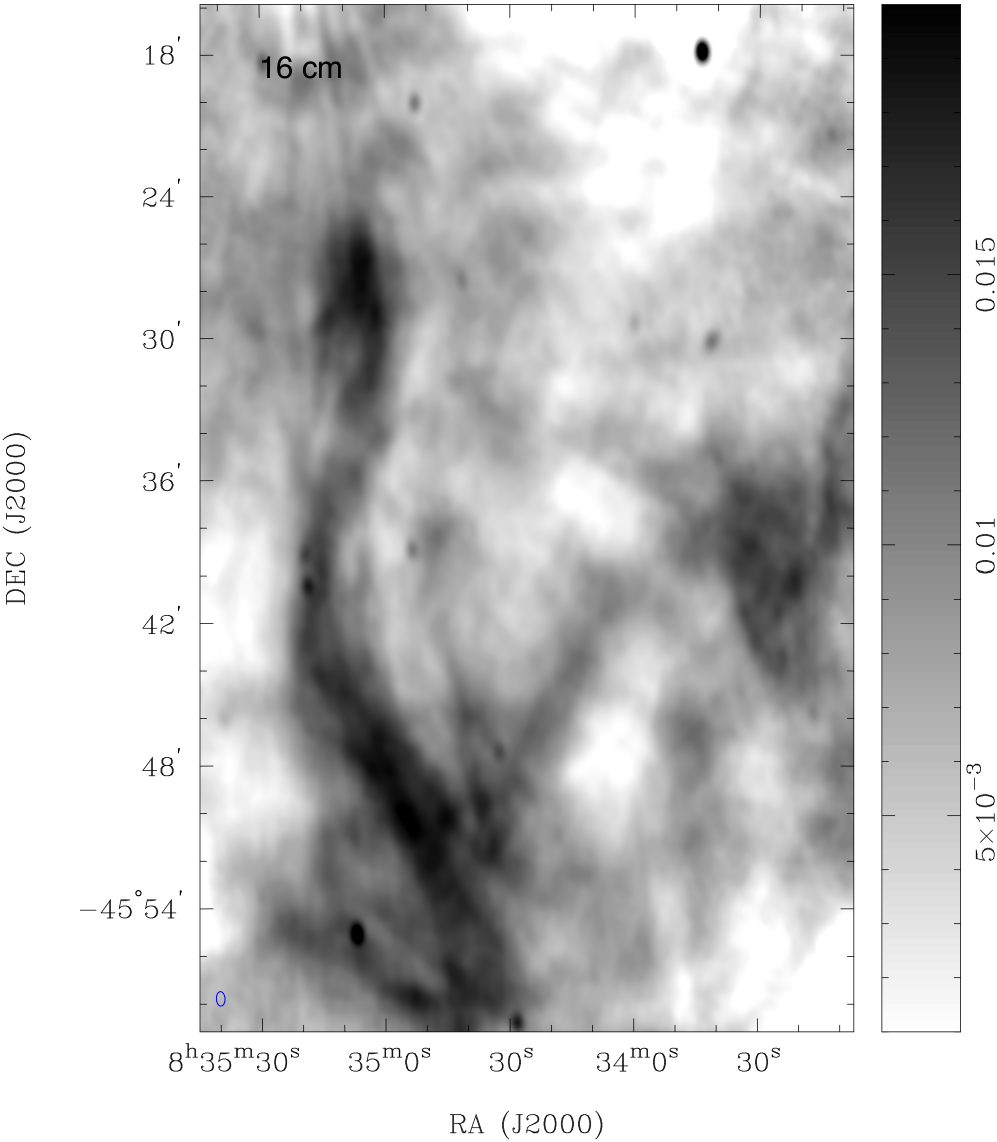}
    \caption{ATCA total intensity maps of Vela X Cocoon at 6 and 16\,cm, with the beams in the bottom left. The gray-scale color bars on the right show the intensity with a unit of Jy\,beam$^{-1}$.   }
    \label{fig:stokesI}
\end{figure*}

We performed new radio observations of the Vela X PWN with ATCA on 2024 November 23, 24, and 25 as project C3579.
All observations used 750\,m array configurations of ATCA (750D) obtain both good resolution and $uv$ coverage. 
The observations on 2024 November 24 and 25 took 3 and 6\,cm data simultaneously with center frequencies at 5500 and 9000\,MHz, and the observation on November 23 took 16\,cm data. 
Unfortunately, the intermediate frequency (IF) 2 receiver of antenna 5 was crashed during all our observations, leading to bad visibility in related baselines, hence data in those IF failed to clearly resolve structures in Vela X Cocoon. Therefore, all IF2 data (i.e., half of 16\,cm and all 3\,cm data) are not used in this work. 
As the observations were carried out after the CABB upgrade \citep{2011MNRAS.416..832W}, each band has bandwidth of 2048\,MHz enabling good sensitivity and also improved coverage in the $uv$ space.
The two observations together cover a $uv$ distance of 0.3 to 95.8\,k$\lambda$ at 6\,cm; for the 16\,cm observation, it covers a $uv$ distance of 0.1 to 45.1\,k$\lambda$. 
The 6 and 16\,cm\ data have total integration times of around 19.6 and 9.6\,hr, respectively, to reach good data qualities. 
1934$-$638 was used as the primary calibrator and 0823$-$500 as the secondary calibrator for all the observations.
All observations have been performed in a mosaicked mode with 25 pointings, so that more interesting regions about the Vela X cocoon region could be covered.
The observation parameters are listed in Table \ref{tab:radio_obs}.

The MIRIAD is a package to reduce, analyze and plot radio interferometric data originally developed for use with ATCA \citep{1995ASPC...77..433S}. We use it for most of the radio analyses in this study. 
Each raw dataset had 40 channels flagged on the edges of the observing band, and we also flagged channels with severe radio frequency interference (RFI) before calibration. 
We used standard procedures within MIRIAD to obtain bandpass, gain, and flux scale solutions from the calibrators. Further RFI flagging was done during calibration as warranted.

For the 6\,cm data, calibration solutions were applied to the target data, we combined the two days of observations and generated mosaicked Stokes \textit{I} dirty image with the task \texttt{invert}, during which process we used multi-frequency synthesis mode and weight the data inversely proportional to noise level, as well as a Briggs weighting with a robust of 1.0 to achieve a better sensitivity.
Maximum entropy routines like \texttt{mosmem} are used to clean the dirty images. 

For the 16\,cm data, we employed the task \texttt{uvmodel} to mitigate contamination of Vela pulsar side lobes, which are on average $\sim$10 times more fluctuating than the theoretical noises.
In this process, we split the data both by pointing and frequency (each with a bandwidth less than 120\,MHz). 
We found pulsar contamination can be substantially suppressed in the cleaned images with narrow bandwidths.   
We then properly extracted a model of pulsar with the task \texttt{clean} and subtract it with the task \texttt{uvmodel} in every individual sub-dataset.
Then the modulated 16\,cm data is used to plot dirty image and then cleaned with mostly identical settings as 6\,cm, though we excluded large baselines with antenna 6 and used a uniform weighting.
More detail about the radio imaging is in Section \ref{sec:result}. 

\section{Results}
\label{sec:result}

\begin{figure*}[]
    \centering
    \includegraphics[width=0.49\linewidth]{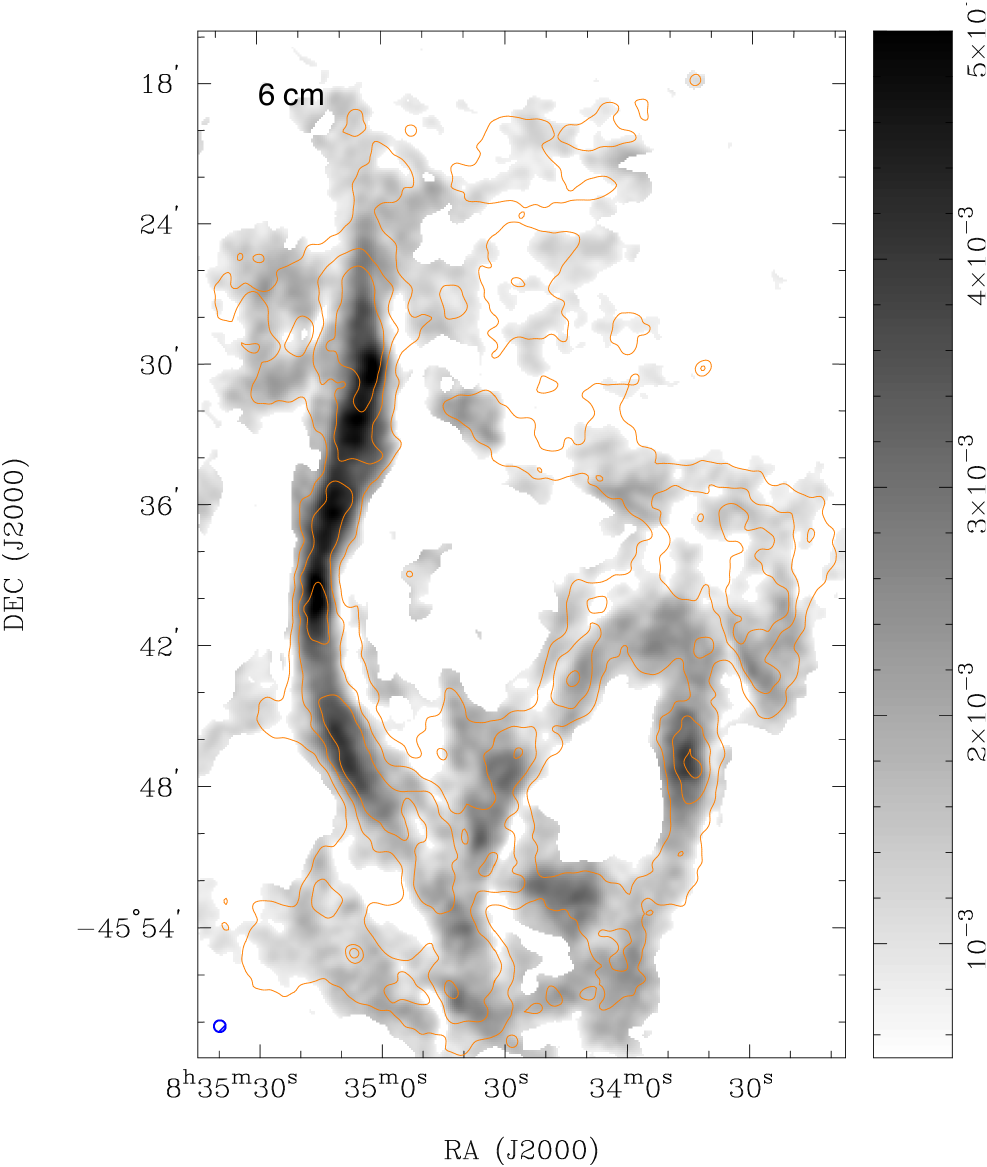}
    \includegraphics[width=0.49\linewidth]{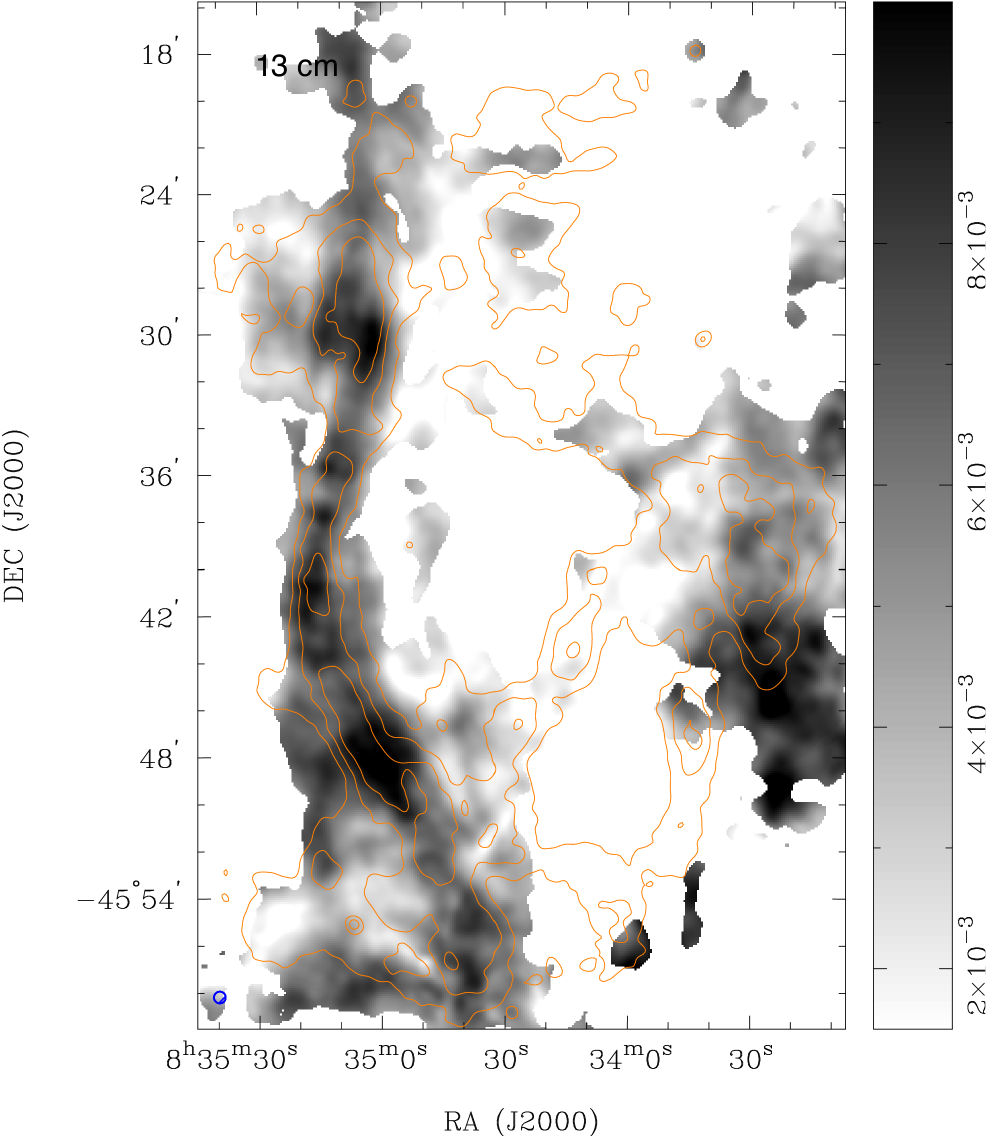}
    \caption{Polarization intensity map of the Vela X Cocoon region at 6\,cm (\textit{left}) and 13\,cm(\textit{right}). The contours are the total intensity map at levels of 2.25, 4.5, 6.75\,mJy\,beam$^{-1}$ at both 6 and 13\,cm. The color bar shows the intensity of polarized emission of the Vela X cocoon region.
    }
    \label{fig:pola}
\end{figure*}

\subsection{Morphology}
Figure \ref{fig:stokesI} shows the ATCA radio images of the  Vela X Cocoon region at 6, and 16\,cm. 
We note that most contamination from the pulsar side lobe has been suppressed in the 16\,cm image, despite of a bit radial artifacts on the top left.
This region is full of filamentary or wisp-like radio structures that are generally aligned with each other at both wavelengths. 
The most significant radio filament (hereafter: the major filament) is right to the south of the Vela pulsar; interestingly, it only becomes bright around 15\arcmin\ south from the pulsar ($\alpha_{2000}\approx\,08^\mathrm{h}\,35^\mathrm{m}\,05^\mathrm{s}, \delta_{2000}\approx-45^{\circ}\,24^{\prime}\,00^{\prime\prime}$).
This filamentary structure has an average thickness of around 2.5\arcmin, and is brightest at the head region, with a peak intensity of 1.8 and 19.3\,mJy\,beam$^{-1}$ at 6 and 16\,cm, respectively. 
Then it extends to the south about 0.5 degree away with a curved morphology (turning $\sim69^\circ$ counter-clockwise overall) making the arrow-like filament end pointing to the southwest.
Besides, there is another filamentary structure in the southwest region, extending northwest from the ``arrow" to around 20\arcmin\ away, with a $\sim$6\arcmin\ large chunk at its end.
Along the filament, there are plenty of small-scale wisp-like structures in the 6\,cm map, as well as many sophisticated wisps around. 
Meanwhile, top-right of both images interestingly shows several east-west parallel filaments with widths of around 2\arcmin.

\subsection{Polarization Map}

We plotted full polarization of 6 and 13\,cm maps\footnote{The 13\,cm map only uses sub-datasets modulated 16\,cm data with center frequencies higher than 2000\,MHz, to prevent significant affects from depolarization at large wavelengths of the 16\,cm dataset.} with a FWHM of 30\arcsec\ for good signal-to-noise (S/N), and then obtained Stokes $Q$ and $U$ models with the task \texttt{pmosmem}. 
Then these restored clean maps produced the polarization intensity (PI), polarization fraction (PF), and $E$-vector position angle (PA) maps using task \texttt{impol}.
In this step, we clipped Q/U map pixels below 0.45 and 1.5\,mJy\,beam$^{-1}$ (3$\sigma$), as well as I map pixels less than 1.5 and 5\,mJy\,beam$^{-1}$(10$\sigma$), at 6 and 13\,cm maps respectively. 
The Ricean bias is also removed in our analysis.

Figure \ref{fig:pola} shows the ATCA image of polarized emission at Vela X Cocoon region, with the contours of total intensities at levels of 2.25, 4.50, 6.75\,mJy\,beam$^{-1}$. 
The polarization intensity map generally follows the morphology in the total intensity map. 
Significant polarized radio emissions are along the major filament, with its peak of 5.2 and 10.3\,mJy\,beam$^{-1}$ in the head region at 6 and 13\,cm, respectively; while other filaments also show some PL emissions, those at 13\,cm are rather weak.
Meanwhile, the highly polarized major filament has average polarization fractions (PFs) of around 55\% at both 6 and 13\,cm, consistent with the previous 3\,cm VLA observation \citep{1995MNRAS.277.1435M}; on the contrary, the average PF of the SW filament is $\sim$50\% at 6\,cm but decreases to $\sim$30\% at 13\,cm. 
It is also notable that emissions are depolarized near the conjunction points at the end of the major filament.
Besides, regions of only faint and chaotic wisps also show quite dim polarization emission.

\subsection{Rotation Measure}

\label{subsec:rm}
\begin{figure}[]
    \centering
\includegraphics[width=1.0\linewidth]{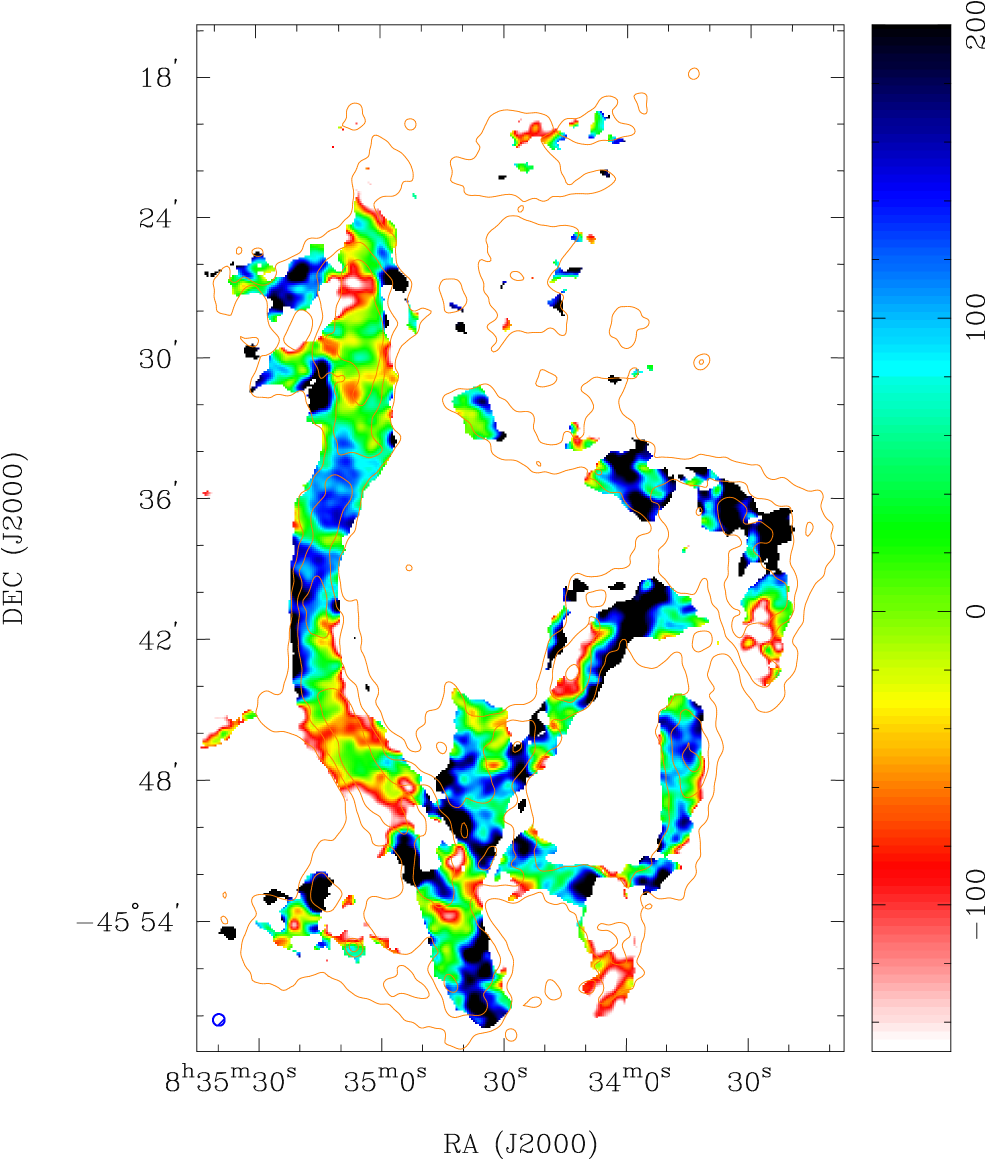}
    \caption{Map of rotation measure in the Vela X Cocoon region. The color bar shows the rotation measure values in the map from $-$150 to $+$100, in units of rad\,m$^{-2}$. 
    The contours show the 6\,cm full-band radio image of cocoon at intensities of 2.25, 4.50, and 6.75\,mJy\,beam$^{-1}$.}
    \label{fig:placeholder}
\end{figure} 

We equally split the 6\,cm data into four frequencies, and used task \texttt{imrm} to obtain the rotation measure (RM) image by linearly fitting produced polarization angle (PA) images. 
Generally, the RM apparently varies in the field of view; taking the major filament as example, the RM value starts from $\sim0$\,rad\,m$^{-2}$ at the head, and gradually increases to over 150\,rad\,m$^{-2}$ around $\sim7$\arcmin\ away in the south.
Interestingly, the RM suddenly jumps to around 0\,rad\,m$^{-2}$ and then back to around 150\,rad\,m$^{-2}$ (within 1\arcmin) near the RM peak ($\delta_{2000}\approx-45^\circ38\arcmin$) where the RM reaches 200\,rad\,m$^{-2}$.
Then it drops to around $\sim-50$\,rad\,m$^{-2}$ to the end of this filament. 
We also note here the RM value of the Vela pulsar is around 30\,rad\,m$^{-2}$ \citep{1977Natur.265..224H},  which is close to the center of RM fluctuation along the major filament, surely consistent with our expectation.
In other regions, the RM values shows rather complicated distribution, from $-$200 to over $+$300\,rad\,m$^{-2}$, fluctuating significantly within small scales.

\subsection{Magnetic Field Structure}
\begin{figure}
    \centering
    \includegraphics[width=1.0\linewidth]{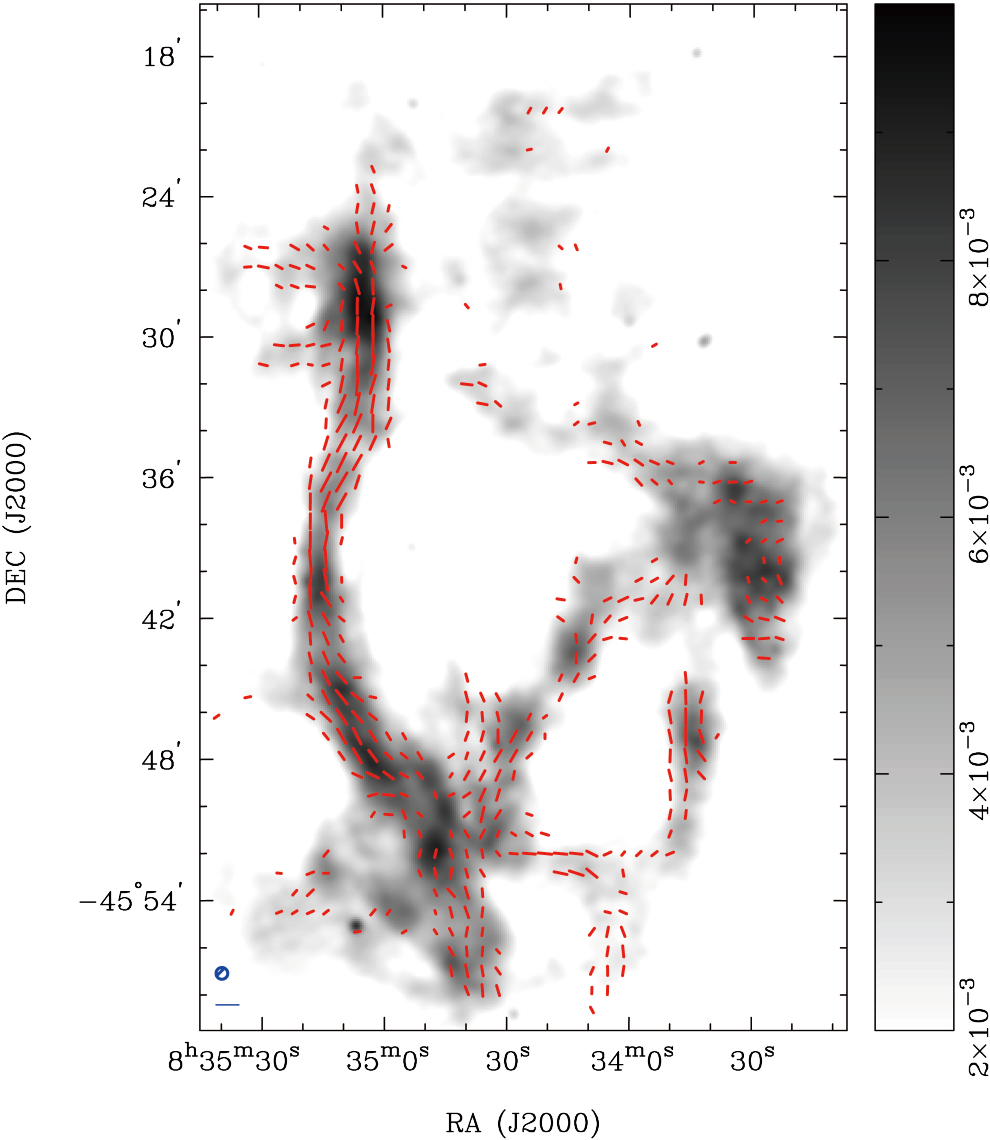}
    \caption{6\,cm ATCA image with $B$-field directions. The vectors show the $B$-field directions in the Vela Cocoon region, the bar at bottom left indicates a polarized intensity scale of 0.5\,mJy\,beam$^{-1}$, which is shown as lengths of $B$-vectors. }
    \label{fig:B-field}
\end{figure}

Figure \ref{fig:B-field} shows the Faraday corrected $B$-vector map of the Vela X cocoon region with the task \texttt{imrm}, as is mentioned in Section \ref{subsec:rm}. 
We discover highly ordered $B$-fields aligning with the elongation of the filaments, for example, $B$-vectors are generally north-south in the major filament and NW-SE in the SW filament. 
However, due to the rather low total intensity from the SNR, $B$-field configurations are not well determined in other regions.

\begin{figure}[ht!]
    \centering
    \includegraphics[width=1.0\linewidth]{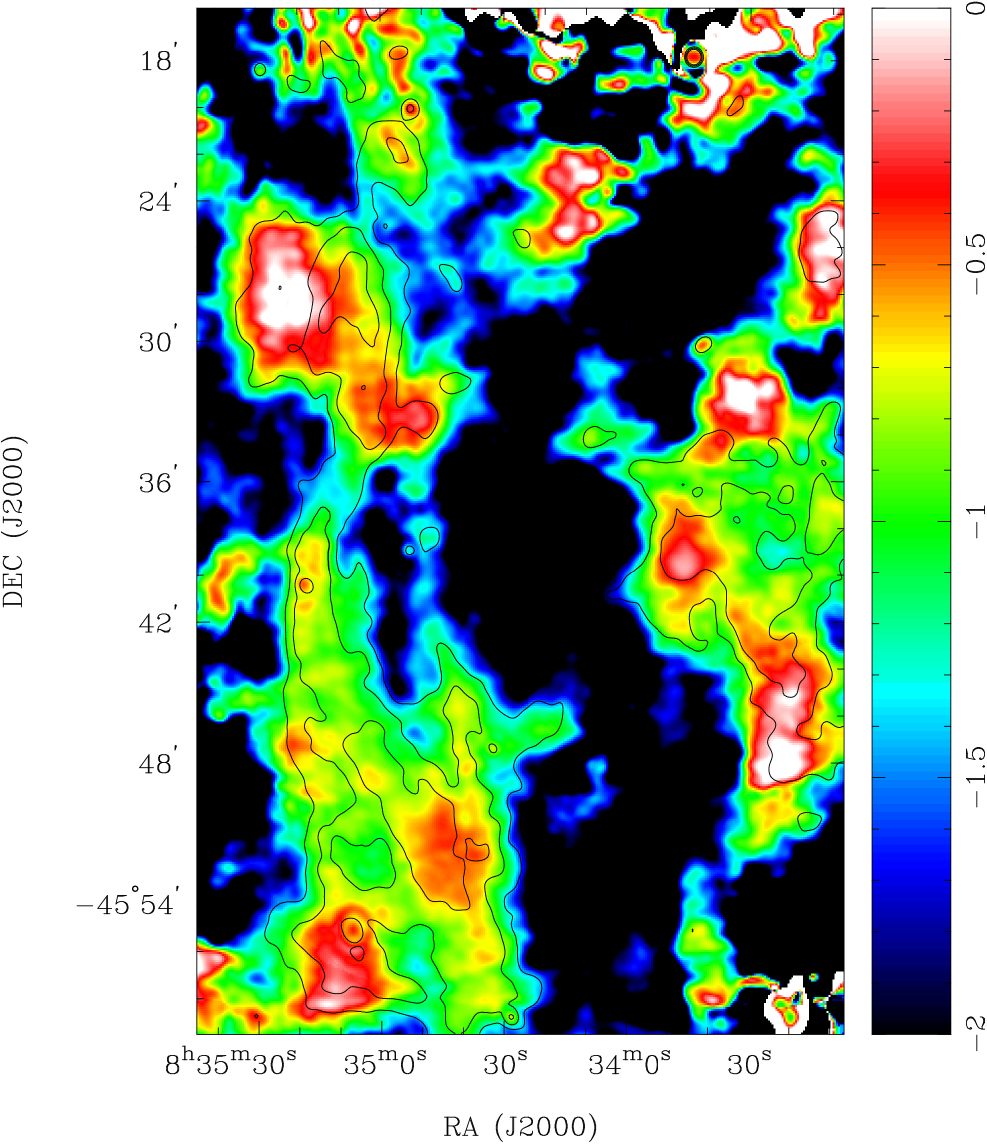}
    \caption{Map of radio spectral index ($\alpha$) in the Vela X Cocoon region from -2 to 0, the black contours show the S-band intensities at levels of 7.5, 11.0, 15.0\,mJy\,beam$^{-1}$.}
    \label{fig:specmap}
\end{figure}

\subsection{Spectral Index Map}

To reveal radio spectral indices of structures in the Cocoon region, we select all sub-datasets below 2000\,MHz to plot an L band map, and all those between 2000 and 3000\,MHz for an S band map, both with a FWHM of 30\arcsec\ and a robust of 0.0. 
We then use cleaned maps to produce the radio spectral index map with the task \texttt{maths}. 

Figure \ref{fig:specmap} shows the distribution of radio spectral indices from $-2$ to 0 in this region. 
The aforementioned radio Cocoon features commonly coincide with the higher spectral index ($\alpha$) regions above $-2$ and some are above $-0.5$ (e.g., in the major filament); on the other side, other regions commonly show much steeper indices below this value, which may reflect the radio spectra of SNR.

\section{Discussions}
\label{sec:discussion}

\begin{figure*}[ht!]
    \centering
    \includegraphics[height=3.in]{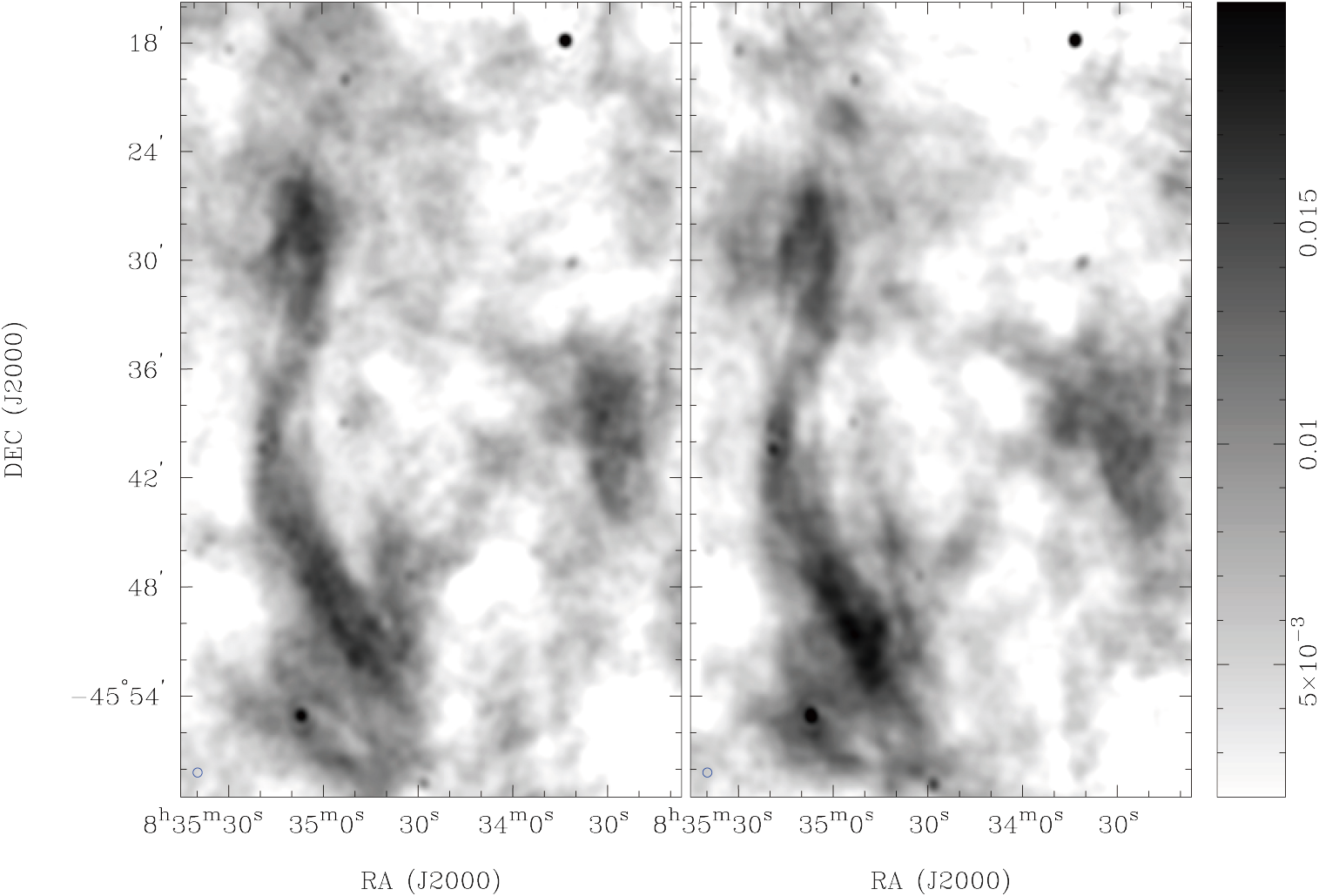}
    \includegraphics[height=3.2in]{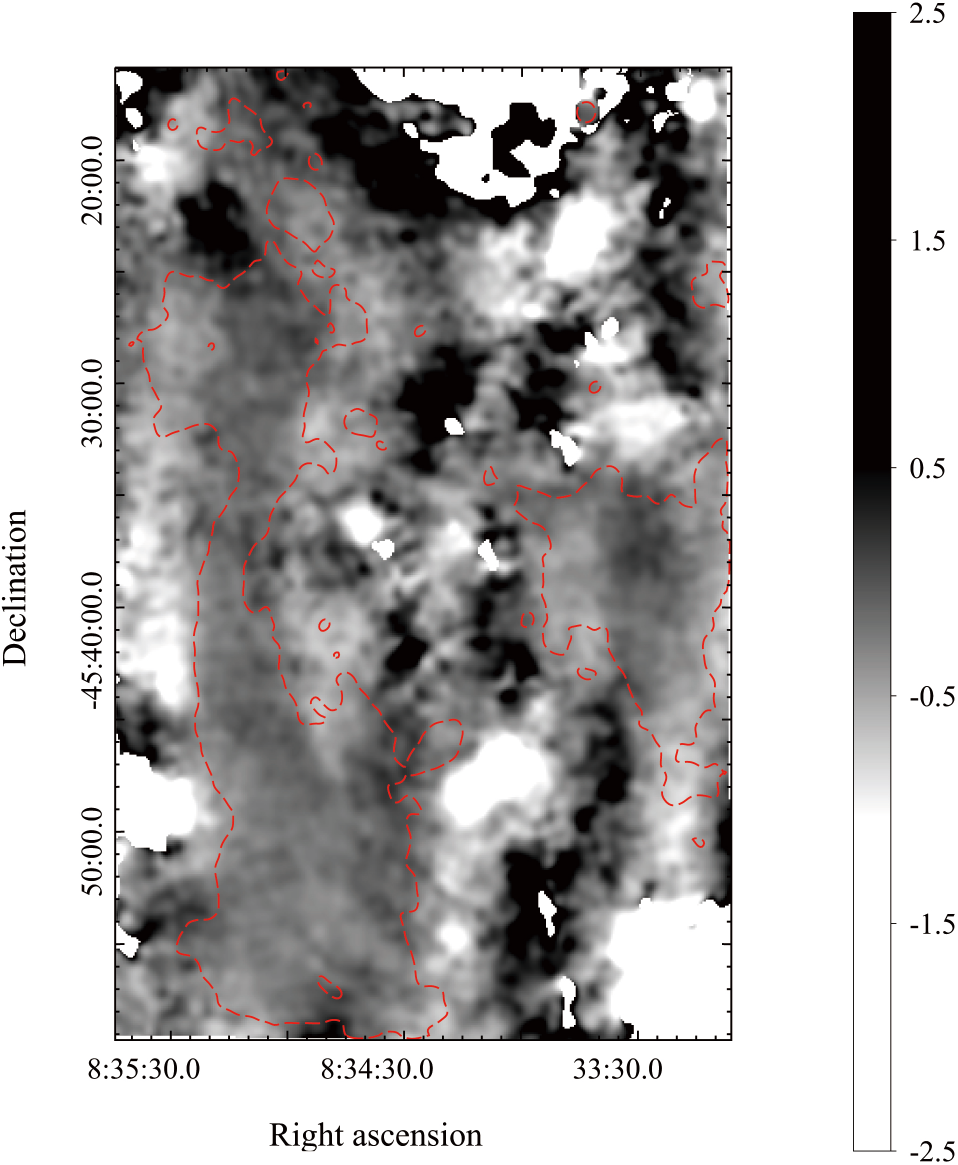}
    \caption{Total intensity images of the S-band ATCA old (\textit{left}) and new (\textit{middle}) observations over 30 years and the relative residual map (\textit{right}) between the two. The red dashed contour indicates to the Cocoon outline in the new observation with intensity at a level of 7.5\,mJy\,beam$^{-1}$ near 2350\,MHz, and both gray scale color bars have a unit of Jy\,beam$^{-1}$.  }
    \label{fig:dynamics}
\end{figure*}

\subsection{Nature of the Radio ``Cocoon"}

As shown in Figure \ref{fig:stokesI}, the radio Vela X cocoon major filament exhibits a curved morphology.
The interesting HESS TeV emissions extending from the pulsar to this region (with a peak near the Cocoon head) imply its capability to generate high energy particles \citep{2006A&A...448L..43A}. 
Observations and hydrodynamic (HD) models have hinted its connections with the RS, which could sweep original PWN matters to far south \citep{2018ApJ...865...86S,2011ApJ...740L..26C,2001PASJ...53.1025M}, while many enigmatic features inside are yet to be properly explained.

One primary question is the origin of the ``Cocoon", especially the major filament.
Apart from the PWN, the SNR is also capable to produce filamentary features, such as G11.2-0.3  \citep{2025ApJ...988..163Z}.
In such a scenario, particles could be accelerated in the turbulent SNR environment and some may also generate synchrotron radiations, whereas radio spectra of SNR features are generally expected to be steeper than PWN. 
In Figure \ref{fig:specmap}, most radio bright regions have spectral indices above $-1$, much harder than other regions dominated purely by SNR materials. 
This suggests that the Cocoon structures may be powered by the Vela pulsar. 
Though many regions are beyond the typical range ($-0.3\lesssim\alpha\lesssim0.0$) for PWN \citep{1978A&A....70..419W}, it can be explained by significant cooling effect after interaction with SNR ejecta. 
It is interesting to note that some high index regions ($\alpha$ close to 0) are not coincident with the most bright regions (e.g., in the major filament). 
Taking the major filament as example, high $\alpha$ features commonly locate near conjunction with other filaments where the $B$-field configurations intriguingly varies significantly and emissions are less polarized.
Regardless of putative missing flux problems, this hints that the RS-PWN interaction could re-accelerate local particles in these regions by introducing turbulent $B$-field environment.

\subsection{Dynamics}
Another question is that the radio and X-ray filaments are distinctive ($\sim4\arcmin$ from each other).
Whether it is due to fast proper motions of filaments (between the radio and X-ray observation) should be precluded in this step.

We here perform the comparison of the previous 2368\,MHz ATCA radio observation in 1996 and our new ATCA observation in 2024. 
For the new 16\,cm data, we selected five sub-bands of dataset (produced in the step introduced in Section \ref{sec:obs_data}) with frequency range most close to 2350\,MHz, for a direct match of flux density with the old observation.
It is notable that we also tried to select similar $uv$ ranges and use a rather large FWHM of 30\arcsec\ for both observations to circumvent the missing flux problems and to optimize S/N.
To detect any displacement of PWN structures, we used MIRIAD task \texttt{imdiff} to reveal the absolute residual map between two observations over 30 years. 
Then we divided the residual map by the new total intensity map ($>$0.8\,mJy\,beam$^{-1}$ for enough S/N) with the task \texttt{maths} to reveal how these features have changed relative to themselves.

Figure \ref{fig:dynamics} reveals the old and new S-band images over 30 years for reference, and the relative residual map between the two on the right.
It is apparent that the general morphologies of the cocoon region are similar over the two epochs, especially for the major filament and the western chunk.
On the contrary, fluctuations are significant in the gap between, generally coinciding with the X-ray bright regions \citep[see Figure 2 in][]{2018ApJ...865...86S}.
It is less likely to extract proper motions in this region, as some detailed features in the old observation (e.g., wisps near the SW filament) have become invisible after 30 years; following observations with shorter intervals can be helpful. 
Nonetheless, this predicts that the X-ray bright region (e.g., the filament) is still highly dynamical, which will not be discussed in the scope of this study.

We note here that a linear fluctuating region extended from the ``gap" intersects with the radio major filament. 
This intersecting position ($\delta_{2000}\mathrm{\approx}-45^\circ36\arcmin$) interestingly coincides with the region of RM and spectral index jumps as demonstrated above. 
These implies its connection with the reverse shock front from the north; under such a hypothesis, the width of such a fluctuating region  ($\gtrsim100\arcsec$) suggest a shock front proper motion of over 3000\,km\,s$^{-1}$. 
Besides, southwest-northeast wisp-like features with widths of $\gtrsim30\arcsec$ are also detected near the end of the major filament, which can also be found in both old and new observations; this suggests a proper motion of no less than 1000\,km\,s$^{-1}$. 
Apart from these, the residual map indeed resolved some large scale features, for example the black chunks on the top, while these features need further accurate estimates with following observations. 

In any case, it can be concluded that the radio cocoon structures are so motionless that cannot move to the X-ray filaments within decades of years (between previous radio and X-ray observations). 
Like Vela X Cocoon, many PWN features (e.g., radio and X-ray jet of the Crab Nebula) are misaligned across the SYN spectrum \citep{2017ApJ...840...82D}. 
If the radio Cocoon is a relic outflow, the X-ray outflow should have switched direction from the head, and it is expected to be highly dynamical in the X-ray region, this is interestingly coinciding with the wisp motions in our result.
Besides, HD simulations \citep{2018ApJ...865...86S,2015ApJ...808..100T} also suggest that the RS will sweep away the original nebula, ram pressure in the trail could then enhance local $B$-field strength and make SYN cooling easier in the radio filaments, for example, highly polarized structures with retained $B$-fields from the RS in the radio Snail Nebula \citep{2016ApJ...820..100M}.
Nonetheless, this need to be supported by following observational evidence (e.g., \emph{eROSITA}), multi-wavelength spectral analysis and further magneto-hydrodynamical simulations.

\begin{figure}[h!]
     \centering
     \includegraphics[width=0.9\linewidth]{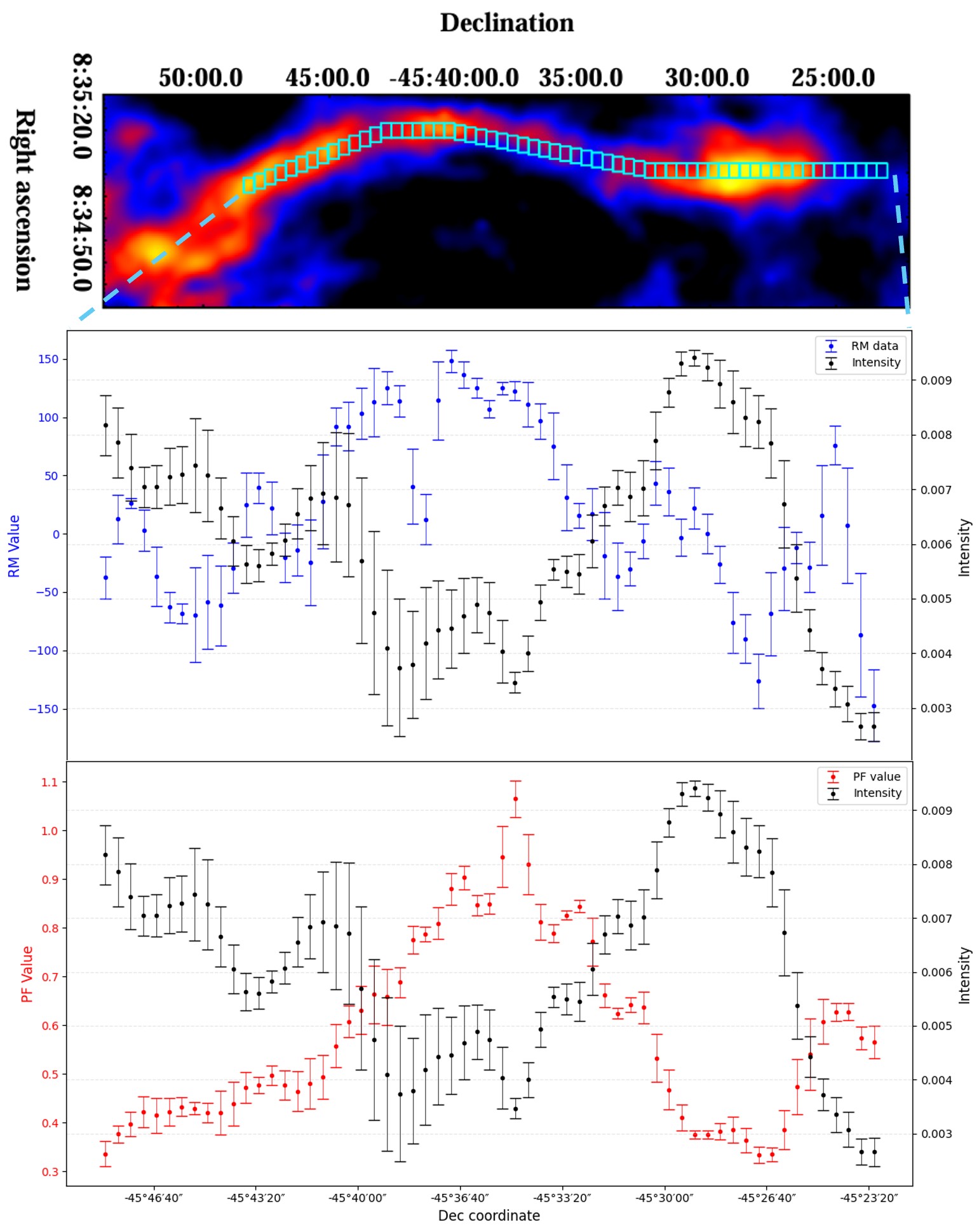}
     \caption{Distributions of rotation measure (blue error bar), polarization fraction (red error bar), and total intensity (black error bar) versus DEC coordinate, extracted from boxes of the top panel.}
     \label{fig:PLstatistics_fila}
\end{figure}

\begin{figure*}[ht!]
     \centering
     \includegraphics[width=0.9\linewidth]{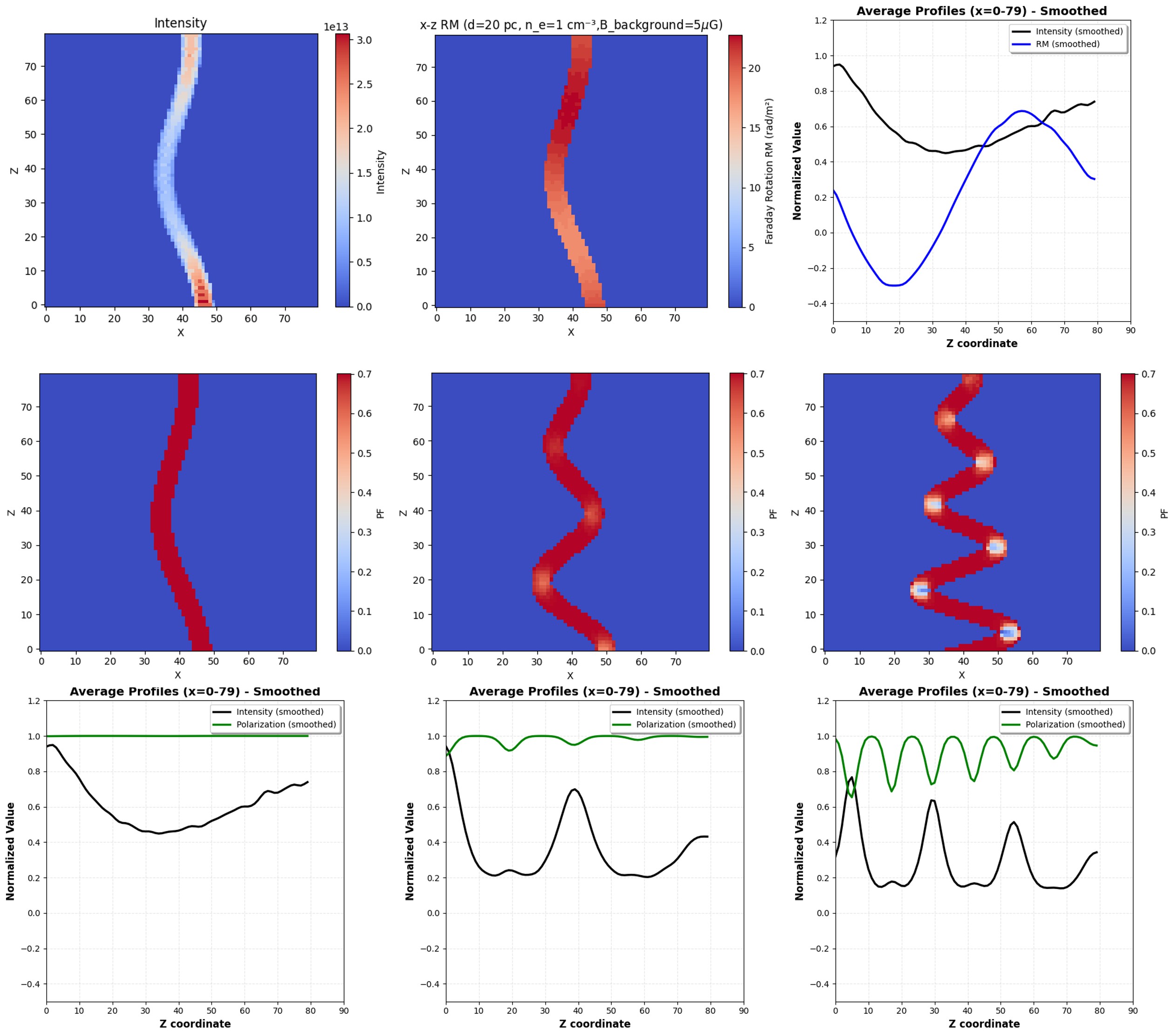}
     \caption{Toy model of total intensity, RM, and PF distributions in a conical solenoid. The top row shows relative intensities across the solenoid with a bulk flow speed of 0.2$c$, the simulated RM map for a solenoid in a uniform $B$-field environment, and comparison of the two with the smoothed intensity and RM distributions in black and blue lines, respectively. 
     The middle row shows the projected polarization fraction geometry of a solenoid with different pitch values of 80, 40, and 30, and the bottom row shows the profiles of smoothed total intensity and PF along the solenoid, with the green line as normalized PF.}
     \label{fig:model_result}
\end{figure*}

\subsection{Modeling of Vela X Cocoon PL and RM Distribution}

Given a distance of $\lesssim300$\,pc, this sample may provide a good probe of high energy PWN charged particle propagation in the SNR environment. 
The curved morphology of the major filament in Vela X Cocoon are also detected in other PWNe (e.g., X-ray helical jet in the Lighthouse Nebula),  suggesting a spiral morphology.
This study estimates distribution of its total intensity ($I$), polarization fraction (PF) and rotation measure (RM) by extracting a set of small rectangular boxes ($35\times 25\arcsec$) over the main body of the major filament from $-45^\circ:22^\prime:57\arcsec$ to $-45^\circ:48^\prime:23\arcsec$ and calculating averages of three aforementioned quantities, and all are shown in Figure \ref{fig:PLstatistics_fila}.  
It is interesting to see that the total intensity ($I$) is interestingly anti-correlated with the other two quantities.  
A simplified toy model is therefore developed to provide a potential physical understanding of these features as below.

We construct a conical magnetic solenoid in a three-dimensional space.
To build such a geometry, we define a helical line on a conical surface of $R =a+b\cdot z$, where $a$ (3 as default) is the initial radius and $b$ (0.6 as default) is factor between cone height and radius. 
Besides, we initially set a pitch along the $z$-axis as 60, to obtain the conical helix line, then all pixels within a distance less than 3 from the helix line is included in the solenoid model.
Within the solenoid, local $B$-fields follows the tangential direction of this solenoid, as is observed in our radio PL observations; while beyond this solenoid, we use a magnetic field $B_f$ in the ambient environment (Vela SNR) with a uniform direction and constant strength all over the space. 
In such a case, we assume uniform $B_f$ in calculation to substitute the RM component of rather turbulent foreground environment; besides, this model also ignores any effects from interstellar medium (ISM) and its $B$-field, due to the rather short distance ($\lesssim300$\,pc).  
The electron number density $n_e$ in the SNR environment is fixed at $1\,\mathrm{cm}^{-3}$, which is not away from the previous hydrodynamical (HD) simulation \citep{2018ApJ...865...86S}.
The model assumes uniform and isotropic internal emissivity within every pixel of this 3D space, with a PL direction perpendicular to local $B$-field in 2D projection. Individual Stokes $I$, $Q$, and $U$ values are then summed up along the line of sight (LoS) to simulate results in our observations.

Firstly, we consider the distribution of the major filament in the total intensity map (hereafter: $I$). 
Relative brightness in our model directly depends on the LoS path length passing through this solenoid, and it is apparent that such circumstance cannot simply lead to large intensity fluctuation as is observed in all frequencies.
We consider the Doppler Boosting effect that putatively exist in PWN structures \citep{2004ApJ...601..479N,2008ApJ...673..411N}, and modulate original intensity $I_0$ of relativistic outflows to $I_m$ according to directions, which can be expressed as
\begin{equation}
    I_m \propto (1-n\cdot\beta)^{-(1+\Gamma)}\,I_0,
\end{equation}
where $\beta=v/c$ is the bulk flow velocity and $n$ is a unit vector pointing to observer in the LoS and $\Gamma$ is the photon index \citep{1987ApJ...319..416P,2004ApJ...601..479N}.  
We here apply such a emissivity modulation by assuming bulk velocities tangential to the geometry of the major filament.
Figure \ref{fig:model_result} (Top-left panel) better reproduces the fluctuating feature of total intensity along the filament, hence requires fast expanding (e.g., $\beta = 0.2$ in our model) PWN outflows in this region.

Secondly, we attempt to investigate the correlation between the RM and total intensity, in which the ratio $B/B_f$
is identified as the dominant parameter. 
Regardless of ISM RM, the RM distribution of the major filament in Figure \ref{fig:PLstatistics_fila} resembles a sine curve. 
Our model considers the RM contributions from foreground medium, which is calculated as
\begin{equation}    
RM \propto \int n_e \cdot \vec{B_\parallel}\mathrm{d}l
\end{equation}
where $\vec{B_\parallel}$ is the $B$-field strength component in the LoS direction. 
Apart from the introduced $B_f$ and $n_e$, we set $d\approx 10$\,pc as the spatial scale of the whole 3D-space to better simulate RM values, and then RM accumulates in the non-uniform foreground paths from the solenoid.
Figure \ref{fig:model_result} (Top-right panels) shows fluctuating RM distribution consistent with the anti-correlation between Stokes $I$ and RM in our observations.
These imply that the RM-intensity morphologies are physically plausible and can be explained with the spatial geometry of the radio filament as investigated in our model. 
Given the inherent uncertainties of such a complex physical system, we refrain from clearly assigning RM distribution by providing a definitive value to the magnetic field strength or the ratio $B/B_f$ in this work.
The radio spectrum might be helpful to estimate the equipartition $B$-field strength in the filaments and environment in the near future, while it is beyond the scope of this work.

Besides, we investigate the 
anti-correlation between $I$ and PF, for which Stokes $I$, $Q$, and $U$ are accumulated along the LoS as illustrated above.
Depolarization effects related to filament geometry are naturally included in our calculations, such as beam depolarization and merging of PL emission at different directions along LoS; however, we here do not consider more complicated scenarios with band-width depolarization and internal RM distribution, which might be possible for future observations with better sensitivity and frequency coverage.
By tuning the cone ratio of the jet, depolarization is present mostly where the projected curve turns. 
In these regions, beam depolarization is dominating by mixing more variable PL emissions along LoS or within a beam.
Results of all these are listed at the middle and bottom rows of Figure \ref{fig:model_result}.
Unfortunately, our results suggest that significant depolarization as  observational results is only shown in cases with a much larger curvature (or smaller pitch) than the major filament in Vela X Cocoon; though considerable Doppler boosting effect is also included, our model also shows that its effect is rather negligible by comparing PF in regions with different bulk flow velocities. 
Therefore our model objects a direct connection between the observed PL distribution and filament geometry, while it is notable that foreground RM is independent in our model and then not precluded.
\section{Conclusion}
\label{sec:conclusion}

In this study, we analyze the radio properties of the Vela X Cocoon region, with the help of our new ATCA observations in 2024. The main points are summarized as follows:
\begin{itemize}
    \item Our high-resolution observations at 6 and 16\,cm reveal the large-scale morphology of curved filaments in this region, as well as chaotic wisp-like features with much smaller scales within or out of filaments. 
   \item The large-scale filamentary features are highly polarized, with the $B$-field tangential to the filaments. RM in the major filament interestingly shows a hump along the major filament while remain fluctuating in other regions. Our current map of radio spectral indices suggests that radio features in the Cocoon region are related to the pulsar rather than purely generated by SNR.
   \item Comparison of old and new ATCA data near 2350\,MHz confirms that radio and X-ray filaments are distinctive, while still implies a highly dynamical environment in this PWN-SNR system. 
   \item We use a toy model to interpret the morphology and PL properties in our observations. Though the total intensity and RM distributions can be explained with a spiral filament with Doppler boosting effect in a chaotic circumstellar medium environment (e.g., SNR), the PL distribution may need further investigations. 
\end{itemize}

It is crucial to further understand some key questions about the Vela X Cocoon like radio and X-ray misalignment with the help of spectral analysis. 
In this case, our observations can work in synergy with multi-wavelength results (e.g., {\emph{Fermi}-LAT}, \emph{XMM-Newton}, and MeerKAT) to constrain the nature of this structure. 
Besides, further investigation might provide important estimate of local $B$-field strength, which works together with our results to build better magneto-hydrodynamical simulations for this system.

\vspace{-1em}
\begin{acknowledgments}
The Australia Telescope Compact Array is part of the Australia Telescope
National Facility (grid.421683.a) which is funded by the Australian Government for operation as a National Facility managed by CSIRO.
C.-Y. N. is supported by a GRF grant of the Hong Kong Government under HKU 17304524. 
This work is also supported by the National Natural Science Foundation of China (NSFC) grants 12261141691.
\end{acknowledgments}

\bibliography{sample7}{}
\bibliographystyle{aasjournalv7}

\end{document}